\documentclass[aps,prl,twocolumn,showpacs,amsmath,amssymb]{revtex4-1}
\usepackage{bookmath} 
\usepackage{graphicx}
\usepackage{amsmath}
\usepackage{color}

\newcommand{\cecoin}{CeCoIn$_5$} 
\def\opp#1{{\overline{ #1}}}

\begin{document}
\title{Enhancement of Electronic Spin Susceptibility in Pauli Limited Unconventional Superconductors}

\author{Benjamin~M.~Rosemeyer}
\author{Anton~B.~Vorontsov}

\affiliation{Department of Physics, Montana State University, Montana 59717, USA}

\date{\today}

\begin{abstract}
We calculate the wave-vector dependent electronic spin susceptibility 
$\chi_{\alpha\beta}(\vq, \vH_0)$ 
of a d-wave superconductor in uniform magnetic field $\vH_0$ with Pauli pair-breaking. 
We find that the transverse component of the susceptibility tensor can be greater than its
normal state value;
the longitudinal component also slightly increases but in a very limited range of $q$s.
We identify several wave vectors $\{\vq_\perp,\vq_\parallel\}$, 
that correspond to the maxima of either $\chi_\perp$ or $\chi_\parallel$. 
We compare our results with available data on the high-field phase in heavy-fermion \cecoin. 
\end{abstract} 



\maketitle

%
Interplay of superconductivity (SC) and magnetism 
has been an active field of research for many years.
Ferromagnetic order 
produces strong uniform internal fields that tend to destroy spin-singlet Cooper pairs. 
Such competition usually results in suppression of one of the orders. \cite{bulaevskii85}
The antiferromagnetic (AFM) order, on the other hand, interferes much less with superconductivity, 
as it gives rise to field oscillations on a short atomic scale, much 
smaller than the Cooper pair size $\xi_0$.\cite{anderson_suhl59} 
Furthermore, in unconventional superconductors under certain conditions 
the superconducting and aniferromagnetic spin-density wave (SDW) orders 
are attractive.\cite{machida87_sdw_hf,*kato87_sdw_hf}

\begin{figure}[t]
\includegraphics[width=1.0\linewidth]{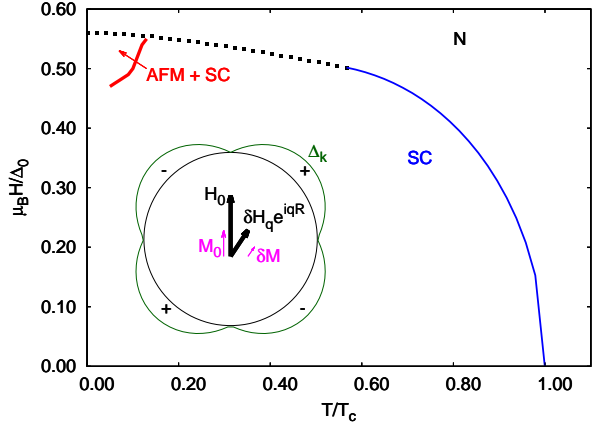}
\caption{ \label{fig:model}
(Color online)
Phase diagram of Pauli-limited superconductor, with the Q-phase\cite{Kenzelmann10_Qphase} sketched. 
We consider circular Fermi surface, and $d$-wave order parameter 
$\Delta_\vk = \Delta_0 (T,H) \sin 2\phi_\vk $. 
The magnetic field has large uniform component $\vH_0$ and spatially varying perturbation $\delta \vH_\vq$
with wave vector $\vq$. 
}
\end{figure}

Recent years have seen another cycle of interest in understanding the details of the SC-SDW interactions 
due to discovery of 
iron-based superconductors\cite{Johnston2010_review} 
and 
Ce-family of heavy-fermion materials\cite{Petrovic01_ce115, Kenzelmann08_Qphase}.
In pnictides the co-existence of the SDW and SC is due to the multi-band nature 
and unconventional order parameter structure. 
The interplay of two orders is a  strong function of the 
Fermi surface (FS) topology.\cite{vor10_sc_sdw,*Fernandes2010_sdw_sc}
In heavy-fermion Pauli-limited \cecoin\ the normal state is non-magnetic but the SDW magnetism (Q-phase)  
appears in the high-field low-temperature part of the phase diagram, through a second-order transition, 
and disappears simultaneously with superconductivity at first-order $H_{c2}$ transition, 
see Fig.~\ref{fig:model}.
\cite{Bianchi2003_fflo, Kenzelmann08_Qphase, Kenzelmann10_Qphase}
The experiments point towards strong AFM fluctuations in the normal state,\cite{Paglione03_qcp115,*Bianchi03_qcp115} 
which, however, are not 
strong enough to produce SDW instability. Nonetheless, these fluctuations can be 
enhanced by doping,\cite{Pham2006_doping115,*Gofryk2012_doping115} 
or possibly by magnetic field, and result in AFM order. 

Following the initial suggestion that the anomlous phase could be  
a non-uniform Fulde-Ferrell-Larkin-Ovchinnikov (FFLO) 
state,\cite{ful64,*lar64}
several theories 
appeared that connected the onset of 
magnetic order to the density of states enhancement by spatially non-uniform SC states, including 
FFLO\cite{yanase11_mag_afm_fflo,Miyake08_sdw_fflo} and vortex cores.\cite{suzuki11_sdw_vortex}.

Another recently proposed explanation of the Q-phase does not require non-uniform SC, 
and is based on the interaction of the uniform superconducting state 
with magnetic field, when Pauli depairing produces favorable conditions for AFM 
instability inside the SC phase.\cite{ikeda10_sc_afm}
The mechanism behind this effect was further revealed in \cite{kato11_sc_afm}, 
which connected the emerging AFM instability 
with the appearance of spin-polarized 
quasiparticle pockets near gap nodes, 
and ``nesting'' of those pockets in momentum space.  

The details of this ``attraction'' between SDW order and Pauli-suppressed SC 
are still not fully uncovered. 
All theories so far assumed only single direction of the SDW ordering vector $\vq$, connecting nodes, 
independent of temperature and the field. 
The size of the SDW phase has not been explicitly connected with the microscopic parameters such as 
size of the SC gap, band width or Fermi energy, and strength of the magnetic interactions. 

In this paper we present a microscopic picture of the 
SDW instability in unconventional $d$-superconductors, 
and find several key features 
consistent with the experiments on \cecoin. 
We calculate the spin susceptibility as a function of magnetization ordering vector $\vq$, 
temperature and field, and determine onset of the magnetic instability in the phase diagram. 
Susceptibility gives detailed information about possible ordering vectors, direction of 
magnetization, and their variations with field and temperature. 
Its magnitude relates the size of the SDW region to magnetic interaction strength, 
SC gap (low) and band (high) energy scales. 
We determine how the ordering vectors at instabilty change 
with field and temperature. 
We find that the mechanism behind enhancement lies not in near-perfect ``nesting'' of new quasiparticle pockets, 
but rather in a combined effect of 
the quasiparticles' dispersion, phase space restrictions and the structure of the order parameter. 
This results in several possible $\vq$ vectors connecting the sharp ends of these pockets. 

%
Our model, $\cH = \cH_{0} + V$, is a mean-field SC Hamiltonian $\cH_0$, 
that includes Zeeman interaction with uniform magnetic field $\vH_0$: 
\be\label{eq:modelH} 
\begin{split}
\cH_{0} = \sum_{\vk \mu} \xi_\vk c^\dag_{\vk \mu} c_{\vk \mu} 
+ \sum_\vk \left( \Delta_\vk c_{\vk \uparrow}^\dag c_{-\vk \downarrow}^\dag + h.c. \right) 
\\
+ \mu_\sm{B} \sum_{\vk \mu \nu} c^\dag_{\vk\mu} \, \vsigma_{\mu\nu} \vH_0 \, c_{\vk\nu}  \,.
\end{split}
\ee
Interaction $V$ is a $\vq$-dependent perturbation 
of the magnetic field $\delta\vH(\vR) = \delta\vH_\vq e^{i\vq\cdot\vR}$, 
$V = \mu_\sm{B} \sum_{\vk \mu \nu} c^\dag_{\vk+\vq \mu}  \, \vsigma_{\mu\nu} \delta\vH_\vq \, c_{\vk \nu}  $, 
where $\mu_\sm{B}$ is the magnetic moment of electron.
The electronic dispersion in the normal state is $\xi_{\vk}=\frac{\vk^2}{2m^*}-\epsilon_f$.  
The resulting magnetization has uniform part and $\vq$-dependent linear response to perturbation:
\be
M_\alpha(\vR) = M_{0\alpha}(\vH_0) + 
\chi_{\alpha\beta} (\vq) \delta H_\beta e^{i\vq \cdot \vR}
\ee
with 
$\vM_0(\vr,t) = \mu_\sm{B} \langle \vS(\vr,t) \rangle_0 $, 
and susceptibility~\cite{mahan}: 
\be
\label{eq:susdef}
\begin{split}
\chi_{\alpha\beta}(\vr, t)= \frac{i \mu_\sm{B}^2}{\hbar} 
\langle [ S_\alpha(\vr,t), S_\beta(0,0) ] \theta(t) \rangle_0 
\\
\chi_{\alpha\beta}(\vq)=  
\int d^3 r e^{-i\vq\vr} \int\limits_{0}^{+\infty} dt \, e^{-0^+ t} \,  
\chi(\vr, t)
\end{split}
\ee
where 
$\vS(\vr,t) = \sum_{\mu \nu} \psi^\dag_\mu(\vr,t) \, \vsigma_{\mu\nu} \, \psi_{\nu}(\vr,t)$,  
$\psi_{\nu}(\vr,t) = \sum_\vk c_{\vk \nu} (t) \varphi_\nu(\vr)$, 
$c_{\vk \nu} (t) = e^{i\cH_0 t} c_{\vk \mu} e ^{-i \cH_0 t}$; 
subscript $0$ indicates the average over ensemble (\ref{eq:modelH}). 

The temperature and magnetic field dependence of the uniform magnetization $\vM_0$
is known, e.g.\cite{Vorontsov06_FLfflo}
and here we discuss the susceptibility $\chi_{\alpha\beta}(\vq)$, since 
it determines the magnetic instability into 
an SDW state, 
and the RKKY-type interaction between localized moments. 
We diagonalize Hamiltonian (\ref{eq:modelH}) by the 
Bogoliubov transformation, 
$c_{\vk \mu} = u_\vk \gamma_{\vk \mu} + (i\sigma_2)_{\mu\nu} v_\vk^* \gamma^\dag_{-\vk \nu} $
with spin-independent coefficients, 
\be
u_{\vk}= \sqrt{\frac{1}{2}\left( 1+\frac{\xi_{\vk}}{\epsilon_\vk}  \right)} \,, \quad 
v_{\vk}= \sgn(\Delta_\vk) \sqrt{\frac{1}{2}\left( 1-\frac{\xi_{\vk}}{\epsilon_\vk} \right)}\,,
\ee
(here $\epsilon_\vk = \sqrt{ \xi_{\vk}^2 + \Delta_{\vk}^2 }$) 
which results in new quasiparticle spectrum 
$
\cH_0 = \sum_{\vk \mu} \epsilon_{\vk \mu} \gamma^\dag_{\vk \mu} \gamma_{\vk \mu} \,,
$ with 
$
\epsilon_{\vk\mu} =\epsilon_\vk \pm \mu_\sm{B} H_0 \,.
$

Using these expressions in (\ref{eq:susdef}), the general formulas for 
longitudinal ($\delta \vM = \chi_\parallel \delta \vH \parallel \vH_0$)
and transverse ($\delta \vM = \chi_\perp \delta \vH \perp \vH_0$)
components of the susceptibility tensor are: 
\begin{widetext}
\begin{subequations}
\label{eq:chi}
\begin{align}
\label{eq:chia}
\chi_{\parallel}( \vq ) = -\mu_\sm{B}^2 \sum\limits_{\vk\mu}  \left\{
\frac{ [ f(\epsilon_{\vk_-\mu}) -f(\epsilon_{\vk_+\mu}) ] ( u_{\vk_+}u_{\vk_-}+v_{\vk_+}v_{\vk_-} )^2} 
     { \epsilon_{\vk_-\mu}-\epsilon_{\vk_+\mu} } 
-\frac{ [ 1-f(\epsilon_{\vk_-\mu})-f(\epsilon_{\vk_+\opp{\mu}}) ] ( u_{\vk_+}v_{\vk_-}-v_{\vk_+}u_{\vk_-} )^2 }
	{ \epsilon_{\vk_-\mu}+\epsilon_{\vk_+\opp{\mu}} }
\right\}
\\
\label{eq:chib}
\chi_{\perp}( \vq ) = -\mu_\sm{B}^2 \sum\limits_{\vk\mu} \left\{
\frac{ [ f(\epsilon_{\vk_-\mu})-f(\epsilon_{\vk_+\opp{\mu}}) ] ( u_{\vk_+}u_{\vk_-}+v_{\vk_+}v_{\vk_-} )^2 }
	{ \epsilon_{\vk_-\mu}-\epsilon_{\vk_+\opp{\mu}} } 
-\frac{ [ 1-f(\epsilon_{\vk_- \mu})-f(\epsilon_{\vk_+\mu}) ] ( u_{\vk_+}v_{\vk_-}-v_{\vk_+}u_{\vk_-} )^2 }
	{\epsilon_{\vk_-\mu}+\epsilon_{\vk_+\mu}} 
	\right\}
\end{align}
\end{subequations} 
\end{widetext}
where 
$f(\epsilon) = [ \exp(\epsilon/T)+1 ]^{-1}$ is the Fermi distribution, 
and momenta are shifted by the magnetization wave vector $\vk_\pm = \vk \pm \vq/2$. 
Notation $\opp{\mu}$ means spin state opposite to $\mu = \pm1$. 

%
In the normal state ($\Delta_\vk = 0$), one obtains the familiar 
Lindhard function,  
\be
\label{eq:chiN}
\begin{split}
\chi^N_{\parallel}(\vq) =- \mu_\sm{B}^2\sum\limits_{\vk\mu} 
	\frac{ f(\xi_{\vk\mu})-f(\xi_{\vk+\vq \mu})}
	{ \xi_{\vk \mu}-\xi_{\vk+\vq \mu} } 
	\\
\chi^N_{\perp}(\vq) =- \mu_\sm{B}^2 \sum\limits_{\vk \mu} 
	\frac{ f(\xi_{\vk\mu})-f(\xi_{\vk+\vq \opp{\mu}}) }
	{ \xi_{\vk \mu}-\xi_{ \vk+\vq \opp{\mu}} } 
\end{split}
\ee
where $\xi_{\vk \mu} = \frac{k^2}{2m^*}-\epsilon_f \pm \mu_\sm{B} H_0$ are electron excitation 
energies in magnetic field. 
At zero temperature the Fermi functions are step-functions, 
and the analytic integration over momenta gives: 
$$
\frac{\chi^N_\parallel(q)}{\chi_0} = 1-\frac12\theta(1-2r_{\uparrow}) \sqrt{1-4 r_\uparrow^2 } 
 -\frac12 \theta(1-2 r_{\downarrow}) \sqrt{1- 4 r_{\downarrow}^2 } \,, 
$$
$$
\frac{\chi^N_\perp(q)}{\chi_0} = 1 - \theta(1-r_{\uparrow}-r_{\downarrow}) 
 \sqrt{ [1- (r_{\uparrow}+r_{\downarrow})^2] [1- (r_{\uparrow}-r_{\downarrow})^2] } \,.
$$
Here $\chi_0 = 2 \mu_\sm{B}^2 N_f$ is the Pauli susceptibility, 
$r_{\uparrow\downarrow} = k_{f\uparrow\downarrow}/q$, 
and $k_{f\uparrow\downarrow} ^2 = k_f^2 ( 1 \mp \mu_\sm{B} H_0/\epsilon_f)$ 
are the Fermi momenta for two spin projections. 
The longitudinal component shows two kinks, at $q=2k_{f\uparrow}$ and $2k_{f\downarrow}$, 
when the Fermi surfaces of up- and down-spins touch at a single point, whereas transverse component involves 
opposite spins which results in only one kink at  
$q=k_{f\uparrow} + k_{f\downarrow}$. 
Generally, the value and behavior of $\chi(q)$ is determined by the properties of the dispersion 
$\xi_\vk$ at hot spots, where $\xi_{\vk+\vq} = -\xi_\vk$. Near those spots both denominator and numerator 
in $\chi$ are close to zero, and the value of the susceptibility is determined by the phase space, 
which is a function of $\vk$-space dimensionality and the shape of the Fermi surface. For example, 
in one dimensional case or for Fermi surfaces with flat parts the susceptibility is logarithmically divergent. 
\cite{roshen83_spin_sus}

%
\begin{figure}[t]
\includegraphics[width=1.05\linewidth]{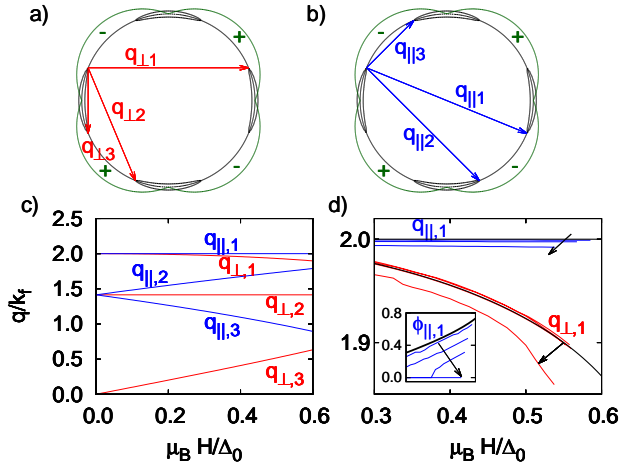}
\caption{ 
	\label{fig:qq} 
	(Color online) 
	(a,b) Magnetic field produces pockets of low energy spin-down excitations near the nodes 
	of the d-wave order parameter. 
	The largest enhancement of $\chi_\perp$ susceptibility 
	occurs when magnetic ordering vector connects ends of quasiparticle pockets with opposite signs of $\Delta_\vk$ (a), 
	or ends with the same sign  $\Delta_\vk$ for $\chi_\parallel$ (b). 
	(c) The magnitude of the ordering vectors as a function of the field at zero 
	temperature, from (a,b).  
	(d) effects of temperature, from $T=0$ to $T=0.3 T_c$ indicated by arrows, 
	on $q_{\parallel,\perp 1}$ vectors, and angle $\phi_{\parallel 1}$ between $\vq_{\parallel 1}$ 
	and x-axis (inset). 
} 
\end{figure}

In the superconducting $d$-wave state we want to find the
maximal values of susceptibility and the corresponding 
magnetization wave vectors. 
Nodal regions of $\Delta_\vk$ in magnetic field host  
spin-down quasiparticles with negative energies, which form new Fermi surface pockets,\cite{kato11_sc_afm} 
and partially destroy superconductivity. 
In the $\vq \to 0$ limit these quasiparticles result in finite 
$\chi_\parallel(0)/\chi_0 \sim \mu_\sm{B} H_0/\Delta_0$. However, the opposite spin coupling in 
the first term of (\ref{eq:chib}) ensures $\chi_\perp (0) = 0$.  
 
Analytic analysis of Eqs.~(\ref{eq:chi}) in general is quite difficult, 
and the result will strongly depend 
on the topology of the Fermi surface, field and temperature. 
However, the important factors to find the vectors $\vq$ that maximize the susceptibility 
can be stated in $T=0$ limit. 
These vectors are shown in Fig.~\ref{fig:qq}(a),(b) for $\chi_\perp$ and $\chi_\parallel$,  
and they connect the sharp ends of the spin-down quasiparticle FS pockets, given by 
$\epsilon_{\vk\downarrow}=\sqrt{\xi_\vk^2 + \Delta_\vk^2}-\mu_\sm{B} H=0$. 
This result is in accord with the enhanced quasiparticle scattering with similar vectors 
observed in \cite{McElroy03_qp_BSCCO}. 
In the vicinity of such common point, 
$\epsilon_{\vk_+\downarrow} \approx \epsilon_{\vk_-\downarrow} \approx 0$ and the 
denominators of the first (second)  term in longitudinal $\chi_\parallel$ 
(transverse $\chi_\perp$) response can be 
expanded as $\vv_+ \delta\vk + \vv_-\delta\vk$. 
The contribution to $\chi$ is greatest 
when the group velocities $\vv_\pm = \grad_\vk \epsilon_{\vk_\pm \downarrow}$ are the 
smallest, \ie near the sharp ends of the banana-like regions, where quasiparticle 
velocity is related to the opening rate of the gap 
$v_\Delta = \partial \sqrt{v_f^2k_\perp^2 + \Delta_0^2 \sin^2 2\phi} / \partial (k_f \phi)
\sim v_f (\Delta_0/\epsilon_f) \ll v_f$. 
The actual magnitude of $\chi$ is determined by the available phase space given by 
complicated FS overlap in 2D $\vk$-plane, 
the distribution functions 
and the superconducting coherence factors in numerators of Eqs.~(\ref{eq:chi}).  
The $\chi_\parallel$'s first term is maximized when 
the magnetization vector $\vq$ connects the same $\Delta_\vk$-sign 
points, making 
$ ( u_{\vk_+}u_{\vk_-}+v_{\vk_+}v_{\vk_-} ) $ 
the most positive and largest with $v_{\vk_+}v_{\vk_-}>0$; 
similarly, largest $\chi_\perp$ is reached when $ ( u_{\vk_+}v_{\vk_-}-v_{\vk_+}u_{\vk_-} ) $
is the most positive. This occurs at vectors, connecting points with 
opposite signs of $\Delta_{\vk_\pm}$. 
The length of the magnetic vectors at $T=0$ 
is shown in Fig.~\ref{fig:qq}(c) as function of magnetic field.

\begin{figure}[t]
\includegraphics[width=1.05\linewidth]{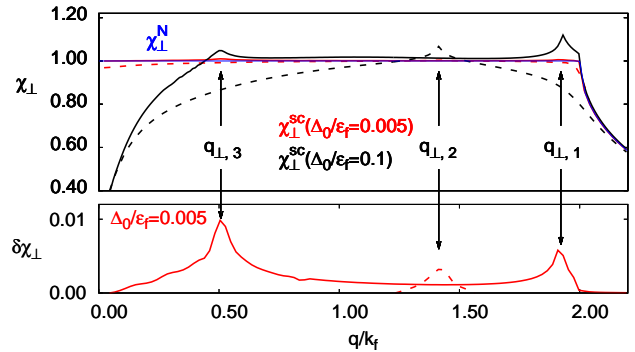}
\caption{
	\label{fig:chi_enh}
(Color online)
The $T=0$ normalized susceptibility in the superconducting and normal
states (blue) as a function of $q$. 
We set $\mu_\sm{B} H = 0.5\Delta_0$, close to
the Pauli limiting field $\mu_\sm{B} H_\sm{P}/\Delta_0 = 0.56$, 
and $\Delta_0/\epsilon_f = 0.005$ (red) or $\Delta_0/\epsilon_f = 0.1$ (black). 
Transverse susceptibility shows enhancement over the normal state $\chi^N(q)$
with 2 peaks at $q_{\perp 1,3}$ for nodal-$\vq$ direction (solid), 
and one peak for $\vq || \vq_{\perp 2}$ (dashed), in accordance with Fig.~\ref{fig:qq}(a,c). 
The lower pane shows zoomed in $\delta\chi_\perp(q) = \chi_\perp^{sc}(q) - \chi_\perp^{N}(q)$ 
for $\Delta_0/\epsilon_f = 0.005$. 
The maximal enhancement $\delta \chi_{\perp}(q)$ occurs at wave vectors 
$\vq_{\perp 1,3}$ and is of the 
order $\delta \chi_\perp/\chi_0 \sim \Delta_0/\epsilon_f$. 
}
 \end{figure}

We confirm this analysis numerically 
and further investigate dependence on the temperature and field, 
on scales $T \sim \mu_\sm{B}H_0 \sim \Delta_0 \ll \epsilon_f$. 
In Fig.~\ref{fig:qq}(d) 
we show the $T$-induced deviations of optimal $\vq_1$ vectors from their $T=0$ values.
At each $T$ and $H_0$ we 
self-consistently compute the amplitude of the gap function $\Delta_\vk = \Delta(T,H) \sin 2\phi_\vk$ 
($\Delta(0,0) = \Delta_0$), which we substitute into Eq.~(\ref{eq:chi}). 
Then we scan over 2D $\vq$ vector to locate  
the maximum of the susceptibility $\chi(\vq_{max})$. 
We find that the ordering vector $\vq_{\perp 1}$ in transverse susceptibility 
follows the zero-T expected pattern, but gets 
reduced with temperature, resulting in smaller overlap of the quasiparticle pockets.
Conversely, for longitudinal component the overlap is increasing with temperature, as 
seen in the inset from the smaller $\phi_{q\parallel}$ angle.  

In Fig.~\ref{fig:chi_enh} we plot $\chi_\perp(q)$ 
in superconducting state at $T=0$ and magnetic field $\mu_\sm{B} H = 0.5\Delta_0$. 
The directions of the ordering vectors $\vq$ are chosen either along the nodal line or along 
$\vq_{\perp 2}$ for this field, see Fig.~\ref{fig:qq}(a).
For the chosen small value of $\Delta_0/\epsilon_f = 0.005$, 
the maximal enhancement of $\chi_\perp$ occurs at the shortest vector $q_3$, 
but we find that the maximum shifts to $q_1$ vector if $\Delta_0/\epsilon_f\sim 0.1$. 

\begin{figure}[t]
\includegraphics[width=1.0\linewidth]{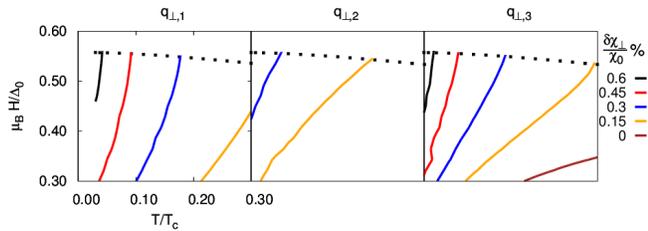}
\caption{ 
	\label{fig:ph.d} 
(Color online) 
Contour lines of maximal enhancement of transverse susceptibility $\chi_\perp$ in the $T$-$H$ 
phase diagram for $\Delta_0 = 0.005\epsilon_f$. Different contours correspond to relative enhancements 
$\delta\chi(q)/\chi_0$, given in percents. 
The three panels correspond to $\vq$-vectors in Fig.~\ref{fig:qq}(a). 
The dotted line is the first order Pauli-limiting phase transition. 
}
\end{figure}

In Fig.~\ref{fig:ph.d} we present the low $T$, high $H$ corner of $T$-$H$ phase diagram 
of a Pauli-limited $d$-wave superconductor, and plot the constant value contours 
of the $\chi_\perp$ peaks, corresponding to different vectors $q_{\perp i}$. 
In this part of the phase diagram 
 $\delta \chi = \chi_\perp^{sc} - \chi_\perp^{N}$ 
becomes positive and progressively larger, while at higher $T$ or lower $H$ 
$\chi^{sc}_\perp(q) < \chi^N$. 
The typical size of the enhancement over the normal state is 
$\delta\chi/\chi_0 \sim \Delta_0/\epsilon_f$. 
The contours of enhanced susceptibility 
$\delta\chi(T,H)$ will determine the boundary of the SDW 
state inside the uniform SC phase, if the magnetic interaction is strong 
enough to cause divergence of 
$\chi^{RPA}(\vq)={\chi(\vq) }/{[1-J(\vq) \chi(\vq)]}$, 
which may happen in case of strong magnetic fluctuations in the normal state, 
$J(\vq) \chi_0 = 1 - O(\Delta_0/\epsilon_f)$. 

We note that the longitudinal susceptibility does not show similar enhancement. 
We find that for $q_{\parallel 1} \sim 2 k_f$ the enhancement 
$\delta \chi_\parallel = \chi_{\parallel}^{sc}(\vq,H_0) - \chi^N_\parallel(\vq,\vH_0)$ can be 
$\sim O(\Delta_0/\epsilon_f)$, but it occurs on the 
background of reduced normal state $\chi_\parallel^N(\vq,H_0)$ and does not lead to 
increase over $\chi_0$. 

These results align very well with the experimental data for \cecoin. 
The general location in the $T$-$H$ phase diagram 
and the shape of the SDW instability, determined by enhancement $\delta \chi_\perp$, 
is consistent with the Q-phase transition, 
and agrees with the conclusions of \cite{kato11_sc_afm}. 
The SC-induced enhancement of $\chi_\perp$, and absence of such in $\chi_\parallel$, explains why the 
Q-phase is observed only when the $H_0$-field is in the $ab$-plane, and the SDW magntization 
is orthogonal to it. 
We find several possible candidates for the SDW ordering vector $\vq$, only one of which 
is probably selected by the magnetic interaction $J(\vq)$. 
The nodal $\vq_{\perp 1}$ is the most likely candidate 
from experimental point of view\cite{Kenzelmann10_Qphase} which sees ordering at $[0.44,0.44]\pi/a$  
(our gap is 45$^\circ$-rotated) 
and it also agrees with the size of the $\alpha$-FS pocket of \cecoin. \cite{suzuki11_sdw_vortex}
The length of this vector drops by about a percent over the $0 - 0.3 T_c$ range, Fig.~\ref{fig:qq}(d), 
and this reduction rate is comparable to change of 0.2\% 
observed in \cite{Kenzelmann10_Qphase} when temperature 
increased  from 60 mK ($0.025 T_c$) to 150 mK ($0.06 T_c$). 
The magnitude of $\chi_\perp$'s enhancement needed to achieve SDW instability 
is also consistent with observed material parameters. 
For the ratio 
$\Delta_0/\epsilon_f \sim 0.6 meV/0.5 eV \sim 0.001$ \cite{Allan2013_stm115,Maehira03_fs115} 
we showed that typical enhancement is of about same size 
$\delta\chi_\perp / \chi_0 \sim \Delta_0/\epsilon_f$  
\ie a fraction of a percent. 
A similar-size enhancement of normal state susceptibility $\chi_0$ can be associated with the FS changes 
induced by Cd-doping CeCo(In$_{1-x}$Cd$_x$)$_5$.\cite{Pham2006_doping115,Hall_2001_FS115,Capan_2010_FS115} 
According to the data,\cite{Capan_2010_FS115} doping of x=0.1 induces AFM state, and corresponds to 5.5\% decrease in FS volume. 
Linear extrapolation of Neel temperature to zero inside SC state gives 4\% minimal doping, 
that would corespond to a 2\% FS decrease, that with a more realistic tight-binding dispersion\cite{kato11_sc_afm} 
corresponds to 3\% increase in $\chi_0$. 
Conversely, applying pressure would increase the FS size, reduce $\chi_0$, and destroy the SDW state.\cite{Pham2006_doping115}

%
In conclusion, we investigated behavior of spin susceptibility in Pauli-limited 
unconventional superconductors. We found that the field-induced nodal quasiparticles, and the sign-changing 
nature of the gap, leads to the enhancement of the transverse susceptibility 
inside the superconducting phase. 
We find several magnetic ordering vectors, connecting sharp (high density of states) ends of the field-induced Fermi pockets. 
The enhancement is of the order $\delta\chi/\chi_0 \sim \Delta_0/\epsilon_f$ 
and is a strong function of temperature and magnetic field; 
it may result in an SDW order formation inside the uniform superconducting phase at low temperatures and 
high fields, whose features are semi-quantitatively consistent with observations in \cecoin.  
To get more detailed agreement with the \cecoin\ data one needs to take into account 
more realistic band structure and 3D topology of the Fermi surface. 

This research was done with NSF support through grant DMR-0954342. 
ABV acknowledges hospitality of Aspen Center for Physics, and discussions with 
I.~Vekhter. 

%

\end{document}